# Reconfigurable Parallel Architecture of High Speed Round Robin Arbiter

Arnab Paul[1], Mamdudul Haque Khan[2], M. Muktadir Rahman[3], Tanvir Zaman Khan[4], Prajoy Podder[5], and Md. Yeasir Akram Khan[6]

[1] Department of ECE at Institute of Engineering & Management (IEM), Kolkata, India
[2,3,4,5] Department of ECE at Khulna University of Engineering & Technology (KUET)
Khulna-9203, Bangladesh
[6] Department of EEE at Rajshahi University of Engineering & Technology, Rajshahi, Bangladesh
arnobpaul11@gmail.com[1], mamdudulhaque@gmail.com[2], muktadir2k9@gmail.com[3], tzkhan19@gmail.com[4], prajoypodder@gmail.com[5], yeasir.akram@gmail.com[6]

*Abstract*— With a view to managing the increasing traffic in computer networks, round robin arbiter has been proposed to work with packet switching system to have increased speed in providing access and scheduling. Round robin arbiter is a doorway to a particular bus based on request along with equal priority and gives turns to devices connected to it in a cyclic order. Considering the rapid growth in computer networking and the emergence of computer automation which will need much more access to the existing limited resources, this paper emphasizes on designing a reconfigurable round robin arbiter over FPGA which takes parallel requests and processes them with high efficiency and less delay than existing designs. Proposed round robin arbiter encounters with 4 to 12 devices. Results show that with 200% increment in the number of connected devices, only 2.69% increment has been found in the delay. With less delay, proposed round robin arbiter exhibits high speed performance with higher traffic, which is a new feature in comparison with the existing designs.

## I. INTRODUCTION

Due to rapid growth in the computer automation sector and computer networking, many devices are seemed to share some common resources. Considering both circuit switching and packet switching technique scenario– they both have some limitations. Circuit switching has trunk limit and packet switching has buffer limit. Now considering the scenario that two users intend to access a single bus at the same time, what will happen? Considering the limitations, a collision may occur and an obvious data loss will happen. Besides there will be unnecessary data flooding and misuse of bandwidth. To compensate this, an access manager block is needed. The access manager block acts as a door way to the bus. It collects requests, processes the requests, makes decisions and open or close the gateway for the particular request. While processing the requests, several considerations may come like equal priority, circular rotation etc. Round Robin Arbitration is a simple time slice scheduling which takes parallel requests and allows each requester a share of the time in accessing a memory or a limited processing resource in a circular order.

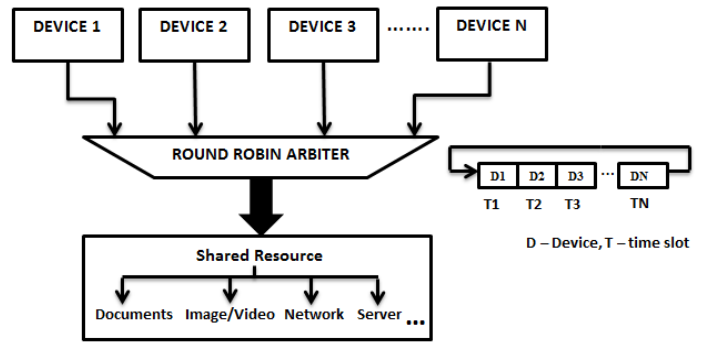

Fig. 1.  Round Robin Arbiter

Its basic algorithm implies that once a requester has been serves he would "go around" to the end of the line and be the last to be served again [11]. Round robin arbiter is used for arbitration for shared resources, networking, queuing systems, and resource allocation [10]. It ensures fairness where none of the requests are suffering from starvation [11], [6].

## II. RELATED WORKS

K. C. Lee proposed a round robin arbiter for high speed application [9]. In his paper, he mainly concentrated on the design procedure and fairness of round robin arbiter. He found good fairness but there was no re-configurability. The delay time made the architecture suitable for lower number of requesting devices but higher number of requesting devices will make that round robin arbiter to work with much delay. There was also port limitation and no RTL schematic shown.

Eung S. Shin, Vincent J. Mooney III and George F. Riley proposed a Round Robin Arbiter Generator (RAG) tool [1]. They also concentrated on design procedure. The delay time seemed to rise almost linearly with the increment of requesting devices which made that architecture suitable for lower number of requesting devices but higher number of requesting devices will make that round robin arbiter to work with much delay. There was no re-configurability. Also there was port limitation and no RTL schematic shown

## III. PROPOSED ROUND ROBIN ARBITER

While designing our proposed round robin arbiter we concentrated on the important facts which were not considered in the previous works. We set 4 goals to achieve

I. Re-configurability

II. Almost Constant Delay Time

III. No Port Limitation

IV. Efficient Space Allocation in FPGA Implementation

Considering these 4 goals, we have designed our round robin arbiter.

## IV. ARBITRATION

An arbitration unit that implements a round robin policy which ensures that each requesting device has an equal chance of accessing a shared resource based upon a current request priority assigned to that requesting device [5]. Figure.2 illustrates the basic arbitration process. The arbitration unit comprises of at least a state block/resister block that includes a plurality of state registers, including a first state register for a first pair of requesting devices, a second state register for a second pair of requesting devices and so on. The register is based on requesting devices.

Arbiter output logic composed of logic gates to determine a request to grant time. The requests are generally tabulated by a fixed order in a multiple bit register with each bit corresponding to the specific request [9]. Arbiter update block takes input from output through a feedback path and tabulates the same way as it comes from arbiter output logic. Arbitration process mainly consists of five stages.

### A. Turn or Token

Each device sharing a common bus has its port connected to the arbiter. The arbiter output is connected to the bus. The round robin arbiter in circular mode with equal priority maintains a time slice [1]. Every device connected to the arbiter is provided with a corresponding time slice by means of a circular rotation. When a machine's turn arrives, the token or turn value of the corresponding machine becomes 1. In other cases, the turn or token value of that machine will be 0.

### B. Request

A round robin arbiter takes parallel requests. When a device needs to be provided with an access to the bus, it has to send a request to the arbiter. A device can independently send request at any time [7]. Upon receiving a request, the request value of the corresponding device becomes 1. In other cases, the request value of that machine will be 0.

### C. Acknowledgement

The arbiter always ands the corresponding token value and the corresponding request value of each device. If the AND output is 1, the arbiter sends the acknowledgement value 1 to the corresponding device. Upon receiving the acknowledgement value 1, that machine realizes that it is now connected to the bus for defined time slice. That is called turn hit. If the AND output is 0, the arbiter sends the acknowledgement value 0 to the corresponding machine.

**Algorithm: Turn hit and Turn miss**

Step 1: Start
Step 2: Declare variables token [N], request [N], ack [N], total_machine and i
Step 3: Read the value of total_machine
Step 4: Initialize variables
    i ← 1
Step 5: Repeat the steps until total_machine=0
    5.1: If i > total_machine
        i ← 1
    Else
        read the value of token[i]
        read the value of request[i]
        ack[i] = token[i] * request[i]
        If ack[i] = 0
            Display Access Denied on Port No. i
        Else
            Display Access Permitted on Port No. i
        i ← i+1
Step 6: Stop

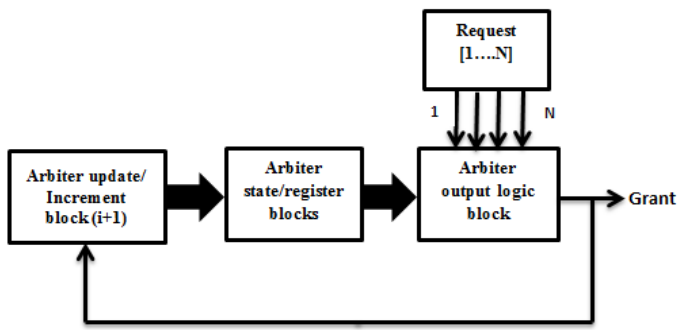

Fig. 2. Arbitration unit block diagram

Fig. 3. Scenario of Arbiter

Upon receiving the acknowledgement value 0, that device realizes that connection to the bus is not possible right now. It is called turn miss [8]. Fig.3 shows the turn hit and turn miss scenario and permitted and denied connection.

*D. Connection Termination*

As for connection termination, there can be two scenarios

I. Time Slice Run out

When the time slice allocated to a particular device runs out, the token or turn value for that device becomes 0 resulting in 0 acknowledgement value and thus connection termination.

II. Termination by request

When a device has been allocated with a time slice, but it completes its task before the time slice runs out, it simply sends the request value 0 to the arbiter resulting in 0 acknowledgement value and thus connection termination. By this, efficient use of time slice can be obtained.

*E. Rounding*

After a connection is terminated, the arbiter makes the token or turn value for the immediate next device 1 keeping other machine's token or turn value 0. This process is maintained in a circular order, one after another device.

While designing a larger machine/device, a quite different form of presentation is used this is called Algorithmic State Machine (ASM) chart. The ASM chart is a type of flow chart used to represent the state transitions and generated outputs.

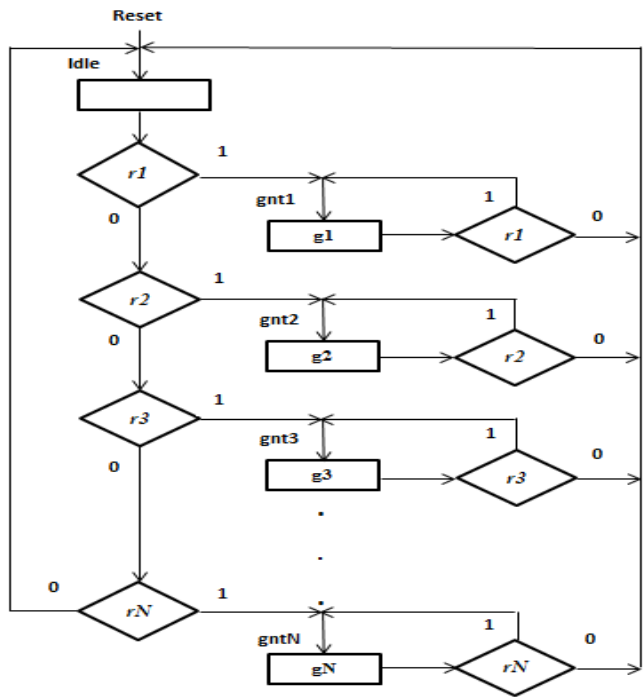

Fig. 4. ASM chart for the arbiter

When all request are zero means $r1, r2… rN=0$ and reset=1, it represents idle condition. When $r1 = 1$, the state changes to gnt1 and output signals g1. The machine stays in state gnt1 as long as $r1 = 1$ and when $r1 = 0$, it changes to state *Idle*. The decision box labeled $r2$ specifies that if $r2 = 1$, then the FSM changes to state gnt2. This decision box can be reached only after first checking the value of $r1$ that is $r1 = 0$. Similarly, the decision box labeled $r3$ can be reached only if both $r1$ and $r2$ have the value 0 [3].

## V. DESIGN PROCEDURE

For a round robin arbiter design, three basic steps must be followed.

I. Modeling
II. Logic Verification
III. Synthesis

First the arbiter is modeled and may be represented in schematic graphs. The designed model is then described in Verilog HDL and the logic can then be verified by using simulation tools or software. When the logic verification is valid, the design can be mapped into a specific process technology, to be synthesized and optimized for area cost and clock frequency [8]. A round-robin arbiter can also be built by means of a shift register and several AND gates. First, it is needed to settle down the issues that the round robin arbiter will have to work with or negotiate. The main theme is the arbiter should keep searching one by one in the cyclic direction [2], [4]. The output notifies if the requested access to the bus is permitted or denied.

The design of an efficient and fast round robin arbiter mostly relies on the capability to search the next requester to grant without losing cycles and with minimal logical stages and skip the non-requesting candidate. It can improve overall system performance. The design requirement of round robin arbiter can be summed up as follows:

1. At each arbitration cycle the round robin arbiter should search for the active requester in a cyclic way.

2. The arbiter does not lose cycle at the end of each round when moving from a grant to the last active requester, back to the first one.

3. Check idle condition (when no request is on, reset =1) and also check grant for different requester in each cycle.

## VI. SYNTHESIZE & IMPLEMENTATION

*A. Circuit Schematic*

Figure. 5 & 6 show the pin diagram and RTL schematic diagram of round robin arbiter. The design of round robin arbiter takes six requests from six parallel connected individual devices. The reset signal clear all registers and control logics. The Clock generate bit rate. The six outputs are for six device input.

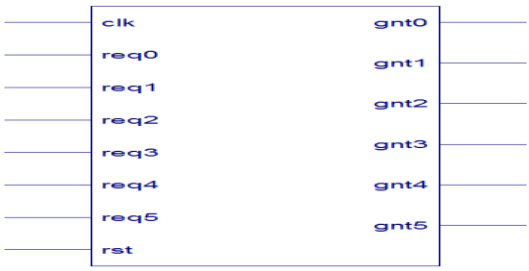

Fig. 5. Pin Diagram of Round Robin Arbiter

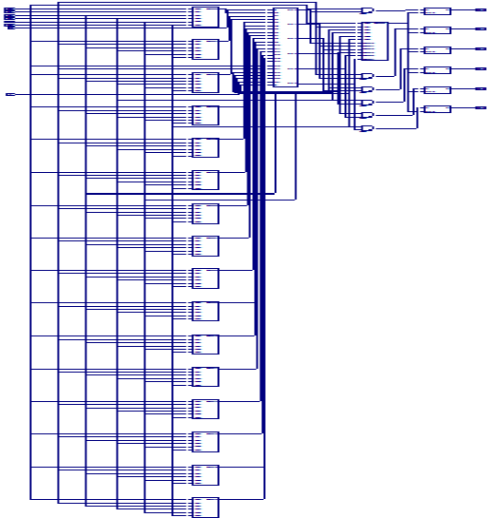

Fig. 6. RTL schematic

We have reconfigured this architecture for up to 12 devices and have taken necessary data. It can be configured up to nth number of parallel inputs.

### B. Simulation Results

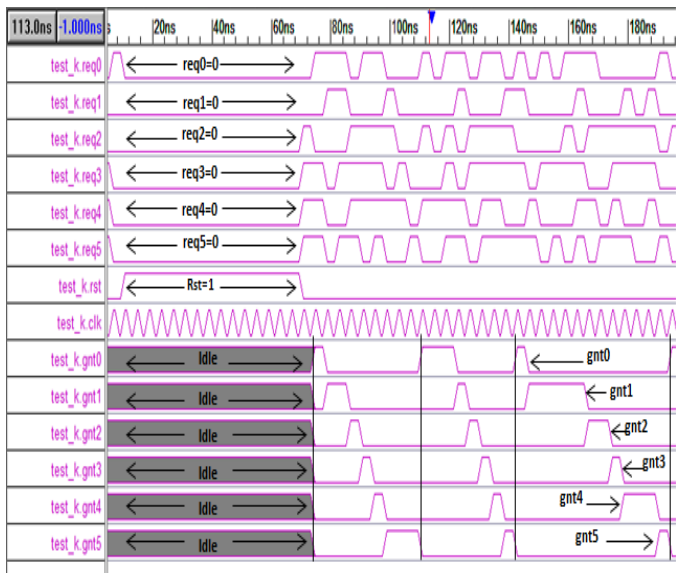

Fig. 7. Simulation of Round Robin Arbiter

The timing diagram achieved from Test-bencher. Figure.7 shows that when the reset is one (reset=1), all requests (from req0 to req5) are zero and output is in idle state (indicated by gray level). When reset goes to zero (reset=0), requests are coming randomly from six devices. All devices (six devices) are active. The grant output is working on the positive edge of clock. Devices have equal chance of getting access to shared resource on priority basis.

### C. FPGA Implementation

FPGAs can be used to implement any logical function and cost-effective. The proposed design is modeled in HDL, using Xilinx 6.3i with SPARTAN 3 family (XC3S1500), FG320 package logically verified and then synthesized in XST synthesis tool. The timing summary, device utilization summary and HDL synthesis report are given in Table I, Table II and Table III respectively. In the timing summary, it is seen that the maximum delay is 3.195ns. From the table II it is seen that only 13flip flop is used among 26624. Table III shows full HDL synthesis report where includes name of components and its quantity.

TABLE I. TIMING SUMMARY

| Parameters | Seconds |
|---|---|
| Minimum period: | 3.195ns |
| Minimum input arrival time before clock: | 5.131ns |
| Maximum output required time after clock: | 5.748ns |
| Maximum delay: | 3.195ns |

TABLE II. DEVICE UTILIZATION SUMMARY

| Name | Used Blocks | Percentages (%) |
|---|---|---|
| Number of Slices | 20 out of 13312 | 0 |
| Number of Slice Flip Flops | 13 out of 26624 (FDR:2 FDRS:4 FDS:1 LD: 6) | 0 |
| Number of 4 input LUTs | 35 out of 26624 | 0 |
| Number of bonded IOBs | 13 out of 221 (IBUF:7 OBUF:8) | 5 |
| Number of GCLKs | 1 out of 8 | 12 |

TABLE III. HDL SYNTHESIS REPORT

| Name | Contents | Quantity |
|---|---|---|
| # FSMs | FSM | 1 |
| # Registers | 1-bit register | 7 |
| # Latches | 1-bit latch | 6 |

TABLE IV. HDL SYNTHESIS REPORT

| Number of devices | Delay(ns) |
|---|---|
| 4 | 3.160 |
| 6 | 3.195 |
| 8 | 3.209 |
| 10 | 3.225 |
| 12 | 3.245 |

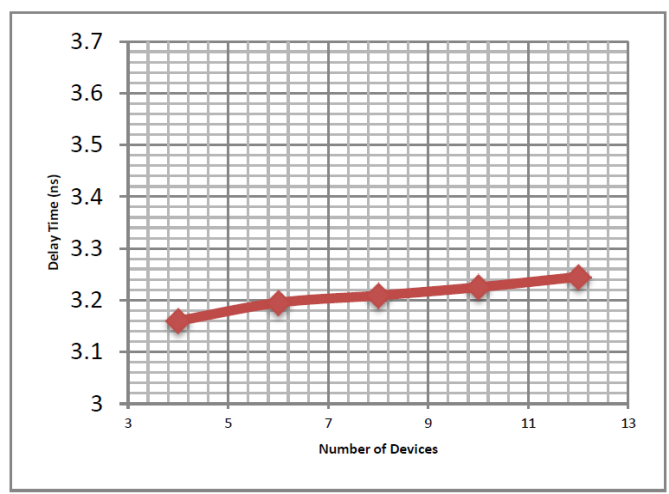

Fig. 8. Number of device vs. Delay time

From the table IV, it has shown that the delay response of designed round robin arbiter with respect to the number of connected/requested active devices is very low. We design the round robin arbiter for several requested devices such as 4 to 12. When number of requested devices are 4, delay is 3.160ns and when requested device quantity are 12, delay is 3.245ns. Only 2.69% increment has found in the delay when requested device have increased from 4 to 12. A graphical figure is shown on fig.8.

## VII. PERFORMANCE ANALYSIS

This reconfigurable round robin arbiter is capable of providing service up to N number of ports maintaining almost same delay with the increase in devices connected to this arbiter. Reconfigurability has made the whole architecture flexible for any further advancement and capacity enrichment. Fig.8 shows that this reconfigurable round robin arbiter maintains almost constant delay which the increment of connected devices which is a unique feature of this round robin arbiter which makes it much suitable for the applications where delay is desired to be maintained the same irrespective of the request load. We have found the switching and synchronizing capability much better. Table II shows that our design comprises less area for implementation ensuring efficient space allocation. Its fairness to requests is also good.

## VIII. CONCLUSIONS

To manage the increased traffic generating due to emergence of computer networking and computer automation, a very fast round robin arbiter design is required to cope up with the speed of high performance buses. In this paper, we have designed a reconfigurable parallel architecture of round robin arbiter for FPGA implementation for high speed application. We have showed that our design has achieved fairness for all requested devices and shows lower delay performance representing only 2.69% increment in delay while been encountered with 200% increased number of devices. This design has been made for smaller space allocation and less complexity. Besides, we are now working to make this design provided with built-in self-test capability. We are trying our best to make this feature added with this design in future.